\pdfoutput=1
\newlength{\absize}
\documentclass[12pt]{article}
\usepackage{graphicx,dsfont,subfig,amsmath,amssymb,setspace,sectsty,hyperref}
\hypersetup{linktocpage}
\renewcommand{\baselinestretch}{1.5}

\sectionfont{\large}
\numberwithin{equation}{section}
\begin{document}
\thispagestyle{empty}
\pagestyle{empty}
\renewcommand{\thefootnote}{\fnsymbol{footnote}}
\newcommand{\starttext}{\newpage\normalsize
\pagestyle{plain}
\setlength{\baselineskip}{3.5ex}\par
\setcounter{footnote}{0}
\renewcommand{\thefootnote}{\arabic{footnote}}}
\newcommand{\preprint}[1]{\begin{flushright}
\setlength{\baselineskip}{3ex}#1\end{flushright}}
\renewcommand{\title}[1]{\begin{center}\Large\bf
#1\end{center}\par}
\renewcommand{\author}[1]{\vspace{2ex}{\normalsize\begin{center}
\setlength{\baselineskip}{3.25ex}#1\par\end{center}}}
\renewcommand{\thanks}[1]{\footnote{#1}}
\renewcommand{\abstract}[1]{\vspace{2ex}\normalsize\begin{center}
\centerline{\bf Abstract}\par\vspace{2ex}\parbox{\absize}{#1
\setlength{\baselineskip}{3.25ex}\par}
\end{center}}
\setcounter{bottomnumber}{2}
\setcounter{topnumber}{3}
\setcounter{totalnumber}{4}
\renewcommand{\bottomfraction}{1}
\renewcommand{\topfraction}{1}
\renewcommand{\textfraction}{0}
\def\draft{
\renewcommand{\label}[1]{{\quad[\sf ##1]}}
\renewcommand{\ref}[1]{{[\sf ##1]}}
\renewenvironment{equation}{$$}{$$}
\renewenvironment{thebibliography}{\section*{References}}{}
\renewcommand{\cite}[1]{{\sf[##1]}}
\renewcommand{\bibitem}[1]{\par\noindent{\sf[##1]}}
}
%\draft
\def\theequation{\thesection.\arabic{equation}}
\preprint{}
\newcommand{\be}{\begin{equation}}
\newcommand{\ee}{\end{equation}}
\newcommand{\ba}{\begin{eqnarray}}
\newcommand{\ea}{\end{eqnarray}}
\newcommand{\bas}{\begin{eqnarray*}}
\newcommand{\eas}{\end{eqnarray*}}
\newcommand{\bc}{\begin{center}}
\newcommand{\ec}{\end{center}}
\newcommand{\nn}{\nonumber}
\newcommand{\comment}[1]{}
\newcommand{\csch}{\mathop{\rm csch\,}}
\newcommand{\sech}{\mathop{\rm sech\,}}
%%%%%%%%%%%%%%%%%%%%%%%%%%%%%%%%%%%%%%%%%%%%%%%%%%
\vspace{4ex}
\title{Glueball Masses from Linearly Confining Supergravity}
\author{Girma Hailu\thanks{hailu@physics.harvard.edu}\\
\vspace{2ex}
\it{Jefferson Physical Laboratory\\
Harvard University\\
Cambridge, MA 02138} }

\abstract{
Mass spectrum of $0^{++}$ glueballs is produced using a dual supergravity theory we proposed for pure $\mathcal{N}=1$ $SU(N)$ gauge theory in four dimensions in the large $N$ limit in the IR. The glueball states are expressed in terms of Whittaker functions. The spectrum is discrete and a function whose roots give the masses is written. The ratios of the masses are independent of the parameters of the theory and comparison to recent non-supersymmetric large $N$ lattice QCD data available for the lowest three states shows agreement to within five percent.
}

\starttext
\newpage
\tableofcontents

\section{\label{sec:intro}Introduction}

One important development in theoretical physics in recent years has been the finding of examples of dual gauge and gravity theories following \cite{Maldacena:1998re, Gubser:1998bc, Witten:1998qj} in which a gauge theory in strongly coupled nonperturbative region has a weakly coupled perturbative description in a gravity theory. The gauge/gravity duality provides the possibility for constructing a calculable theory of the strong interactions at low energies.
Understanding the nonperturbative low-energy physics of quantum chromodynamics (QCD) is crucial, for instance, to formulating a quantitative description of the mass spectra of bound states of quarks and gluons.

We proposed in \cite{Hailu:2011pn, Hailu:2011kp} a gravity dual to pure $\mathcal{N}=1$ $SU(N)$ gauge theory in four dimensions in the large $N$ limit in the IR which reproduces the renormalization group flow and the pattern of chiral symmetry breaking of the gauge theory and produces linear confinement of quarks within relativistic quantum theory of ten-dimensional superstrings for the first time.

In this note we study fluctuations of the dilaton on the gravity background to calculate mass spectrum of $J^{PC}=0^{++}$ glueballs, spin-0 bound states of gluons which are even under parity and charge conjugation. The masses are given by
\be
m_n=\frac{g_sN}{2\pi r_s}\lambda_n,\quad n=1,2,\cdots,\label{mn-1}
\ee
where $\frac{g_sN}{2\pi r_s}$ is a combination of the parameters of the theory and $\lambda_n$ are the roots of the function
\be
f(\lambda)=M_{\frac{\lambda}{4}+1,-\frac{1}{4}}(\lambda)+\frac{\lambda-1}{\lambda+1}\,
   M_{\frac{\lambda}{4},-\frac{1}{4}}(\lambda)\,,\label{lambdaroots-1}
\ee
where $M_{\frac{\lambda}{4}+1,-\frac{1}{4}}(\lambda)$ and $M_{\frac{\lambda}{4},-\frac{1}{4}}(\lambda)$ are
Whittaker functions of the first kind.

The spectrum is discrete.
The ratios of the masses are independent of the parameters and serve as predictions for $0^{++}$ glueball masses in $\mathcal{N}=1$ $SU(N)$ gauge theory in the large $N$ limit. Comparison of the ratios to recent large $N$ non-supersymmetric lattice QCD data available for the lowest three states shows agreement to within five percent.

\section{Supergravity background\label{sec:sugrab}}

In this section we briefly summarize features of the supergravity background in \cite{Hailu:2011pn, Hailu:2011kp}. The starting point of the construction is the supergravity background of unwrapped and flat $N$ D7-branes on $\mathds{C}^1/Z_2$ with running axion and dilaton fields and the eight-dimensional $\mathcal{N}=1$ $SU(N)$ gauge theory that lives on the branes. The background geometry of the eight-dimensional gauge theory is noncompact with a conic singularity. The D7-branes are then wrapped over a 4-cycle on $\mathds{C}^1/Z_2\times \mathds{T}^2/Z_2 \times \mathds{T}^2/Z_2$ to construct $\mathcal{N}=1$ $SU(N)$ gauge theory in four dimensions. The wrapped D7-branes turn on all $F_1$, $F_3$, $H_3$, and $F_5$ fluxes of type IIB theory and induce torsion. $\mathcal{N}=1$ supersymmetry in four dimensions is preserved and the background is stable by a balance between fluxes and torsion. The conic singularity in the background of the eight-dimensional gauge theory is smoothed out by the fluxes in the background of the four-dimensional gauge theory on the wrapped branes.  The supergravity solutions are obtained by solving the supersymmetry equations in \cite{Hailu:2007ae} which were generalizations of the equations in \cite{Grana:2004bg,Butti:2004pk}. The solutions to the supersymmetry equations with the Bianchi identities imposed on them then automatically solve the bosonic supergravity equations of motion.

The extra six-dimensional space of the ten-dimensional spacetime is parameterized by one radial coordinate $r$ with range $r_s\le r\le r_s\,e^{\frac{2\pi}{3g_sN}}$ and five angles $\psi$, $\varphi_1$, $\varphi_2$, $\varphi_3$, and $\varphi_4$, each having the range between 0 and $2\pi$; $r$ and $\psi$ are coordinates on $\mathds{C}^1/{Z_2}\sim \mathds{R}^1\times\mathds{S}^1$ and the remaining angles are coordinates on $\mathds{T}^2/Z_2\times \mathds{T}^2/Z_2$. We call $r=r_s$ the IR boundary and $r=r_s\,e^{\frac{2\pi}{3g_sN}}$ the UV boundary; $g_s N$ is the 't Hooft coupling at the UV boundary.
The $N$ D7-branes are wrapped over $\mathds{T}^2/Z_2\times \mathds{T}^2/Z_2$ at $r=r_s$.
The background metric is
\ba
&{d{s^0}_{10}}^2=\cosh u\,dx_{1,3}^2
+ r_s^2\,\sech u\,(d\rho^2+d\psi^2&\nn\\&+d\varphi_1^2+d\varphi_2^2
+d\varphi_3^2+d\varphi_4^2)\Bigr),\label{metric-3n}&
\ea
where
\be
\sech u= \sqrt{1-(\frac{g_sN}{2\pi}\rho)^2},\quad \rho\equiv 3 \ln (\frac{r}{r_s}),\label{hn-1}
\ee
and $dx_{1,3}^2$ is the metric on flat four-dimensional spacetime $\mathds{R}^{1,3}$ with coordinate $x^\mu$. We use uppercase indices $M,N,\cdots$ to denote the coordinates of the ten-dimensional spacetime. A superscript of $0$ is put to denote background metric and fluxes excluding fluctuations.

The dilaton and the background fluxes are
\ba
&e^{\Phi^0}=g_s,\qquad F_1^0=\frac{N}{2\pi}d\psi,&\nn\\ &F_3^0=\frac{N}{2\pi}\,r_s^2\,\tanh u \, \left(d\psi\wedge d\varphi_1\wedge  d\varphi_2 +d\psi\wedge  d\varphi_3\wedge  d\varphi_4\right),&\nn\\ &H_3^0=\frac{g_sN}{2\pi}\,r_s^2\,\sqrt{1+2\cosh 2u}\left(dt\wedge d\varphi_1\wedge  d\varphi_2+dt\wedge d\varphi_3\wedge  d\varphi_4\right), &\nn\\
&\tilde{F}_5^0=(1+\star_{10})F^0_5=\frac{N}{2\pi}\,r_s^4\,\left(1+\tanh^2u\right) \, d\psi\wedge d\varphi_1\wedge  d\varphi_2\wedge  d\varphi_3\wedge  d\varphi_4&\nn\\ &-\frac{N}{2\pi}\,\sqrt{1+2\cosh 2u}\,\cosh 2u\, \cosh^2u \, d^4x \wedge dt,
\qquad \,& \label{F5sol-1bbn}
\ea
where
\be
dt=\frac{d\rho}{\sqrt{1+2\sech 2u}}.\label{rut-1n}
\ee

The background metric and fluxes above solve all the bosonic supergravity equations given by
\be
\frac{1}{\sqrt{-G}}\,\partial_M\left(\sqrt{-G}\,e^{-2\Phi}G^{MN}
\partial_N\Phi\right)=
-\frac{1}{8}F_1^2-\frac{1}{96}F_3^2+\frac{1}{96}e^{-2\Phi}H_3^2,
\label{phi-1n}
\ee
\ba R_{MN}=& &-4\,\partial_M\Phi \partial_N\Phi +\frac{1}{2}e^{2\Phi}\partial_M C_0\partial_N C_0+\frac{1}{4}(H_3)_{MOP}(H_3)_{N}^{\hspace{3mm}OP}\nn\\ & &+\frac{1}{4}e^{2\Phi}(F_3)_{MOP} (F_3)_{N}^{\hspace{3mm}OP}+\frac{1}{96}e^{2\Phi}(\tilde{F}_5)_{MOPQR}
(\tilde{F}_5)_{N}
^{\hspace{3mm}OPQR}\nn\\ & &
-G_{MN}\,(\frac{1}{48}H_3^2+\frac{1}{48}e^{2\Phi}F_3^2
+\frac{1}{960}e^{2\Phi}\tilde{F}_5^2),
\label{RMN-1n}
\ea
\be dF_3=-F_1\wedge H_3,\qquad  d\tilde{F}_5=H_3\wedge F_3,\label{Bianchi-2a}\ee
\ba
&d\star_{10}\left(e^{2\Phi}F_1\right)=-e^{2\Phi}H_3\wedge \star_{10}F_3,\qquad d\star_{10}\left(e^{2\Phi}F_3\right)=F_5\wedge H_3,&\nn\\ &d\star_{10}\left(H_3-e^{2\Phi}C_0 F_3\right)=-e^{2\Phi}F_5\wedge F_3.&\label{bosonic-2n}
\ea

The radial coordinate $r$ is mapped to the scale $\Lambda$ of the gauge theory as  $\Lambda/\Lambda_s\sim r/r_s$, where $\Lambda_s$ is the nonperturbative scale at which the gauge coupling formally diverges.
The coordinate $\psi$ is mapped to the Yang-Mills angle $\Theta$ in the gauge theory as $\Theta=N\psi$.
The gauge coupling of the of the four-dimensional gauge theory is related to the gauge coupling of the eight-dimensional gauge theory and the volume of the 4-cycle wrapped by the D7-branes, and it runs by inheriting the running of the dilaton from the background of the eight-dimensional gauge theory. The axion potential and the dilaton are given by $C_0=\frac{N}{2\pi}\psi$ and $e^{-\Phi}=\frac{N}{2\pi}\,\rho$ in the background of the eight-dimensional gauge theory.
The background geometry of the four-dimensional gauge theory is compact and conformally Calabi-Yau.
The 't Hooft coupling of the four-dimensional gauge theory is given by $\frac{g_4^2N}{4\pi}=\frac{2\pi}{\rho}$, which is large in the IR and equals $g_sN=e^{\Phi^0}N$ at the UV boundary.
The curvature is smallest in the IR region where the gauge theory is strongly coupled and a dual gravity description is useful. The internal space normal to the wrapped D7-branes at the IR boundary is $\mathds{S}^1$ whose radius is set by the nonperturbative scale of the gauge theory independent of the 't Hooft coupling and spacetime is four-dimensional at the UV boundary. The potential energy of a quark-antiquark pair located at the UV boundary increases linearly with their spatial separation in four dimensions.

\section{$0^{++}$ glueball masses\label{sec:gbm-1}}

We want to study $0^{++}$ glueballs in four dimensions. The masses of $0^{++}$ glueballs in the gauge theory are related to the two point correlation function of the square of the gluon field strength tensor $\langle \mathrm{tr}F^2\,\mathrm{tr}F^2\rangle$. Following a prescription of the gauge/gravity duality \cite{Maldacena:1998re, Gubser:1998bc, Witten:1998qj}, the glueball states are mapped to supergravity states that couple to the operator $\mathrm{tr}F^2$. Because the dilaton couples to $\mathrm{tr}F^2$, the $0^{++}$ glueball masses can be computed by studying linear fluctuations of the dilaton \cite{Witten:1998zw, Csaki:1998qr}. To write appropriate equation of motion, we start with $\Phi=\Phi^0\,+\,\delta \Phi$, $F_1=F_1^0\,+\,\delta F_1$, $F_3=F_3^0\,+\,\delta F_3$, $H_3=H_3^0\,+\,\delta H_3$, $F_5=F_5^0\,+\,\delta F_5$, and $G_{MN}=G_{MN}^0\,+\,\delta G_{MN}$, where the second terms are the fluctuations.
Because the dilaton field in the background of the four-dimensional gauge theory is constant, $e^{\Phi^0}=g_s$, the left hand side of (\ref{phi-1n}) contains  only fluctuations of the dilaton but not of the metric at linear order and we write (\ref{phi-1n}) as
\be
\partial_M(\sqrt{-G^0}\,{G^0}^{MN}\partial_N\delta\Phi)=0,\label{phi-eqn-1}
\ee
where linear order fluctuations in the fluxes and the metric need to satisfy  $-\frac{1}{{2H_3^0}^2}(12\,e^{2\Phi^0}\delta(F_1^2)
+e^{2\Phi^0}\delta(F_3^2)-\delta(H_3^2))=
\delta\Phi$ in order for the background with the linear fluctuations to preserve supersymmetry.
The equations for the fluctuations of the metric and the fluxes would then involve $\delta\Phi$. Our interest here is to study fluctuations of the dilaton governed by the decoupled equation given by (\ref{phi-eqn-1}).

To find the mass measured by a four-dimensional observer, we make a plane wave expansion of the form $\delta\Phi(x,\rho)=\phi(\rho)\,e^{ik.x}$ and the glueball mass $m$ is given by $m^2=-k^2$.
Using the metric given by (\ref{metric-3n}) in (\ref{phi-eqn-1}),
\be
\phi''(y)+\lambda^2 (1-y^2)\phi(y)=0,\label{phi-eqn-2}
\ee
where
\be
\lambda \equiv \frac{2\pi r_s}{g_s N}\,m,\qquad y\equiv \frac{g_s N}{2\pi}\rho,\label{lambda-1}
\ee
and $0\le y\le 1$. But (\ref{phi-eqn-2}) can be written as a Whittaker equation
\be
w''(z)+\left(-\frac{1}{4}+\frac{\lambda}{4z}+\frac{3}{16z^2}\right)w(z)=0,\label{whittaker-1}
\ee
where
\be
w\equiv\sqrt{y}\,\phi, \qquad z\equiv \lambda y^2.\label{fx-1}
\ee
The general solution to (\ref{phi-eqn-2}) can then be written as
\be
\phi(y)=\frac{1}{\sqrt{y}}\left(C\, M_{\frac{\lambda}{4},\frac{1}{4}}(\lambda y^2)+D\, W_{\frac{\lambda}{4},\frac{1}{4}}(\lambda y^2)\right),\label{phi-sol-1}
\ee
where $M_{\frac{\lambda}{4},\frac{1}{4}}(\lambda y^2)$ and + $W_{\frac{\lambda}{4},\frac{1}{4}}(\lambda y^2)$ are respectively Whittaker functions of the first kind and the second kind and $C$ and $D$ are constants.

Now we need to impose appropriate boundary conditions at $y=0$ and $y=1$. We demand that physically acceptable modes of fluctuation be smooth at both boundaries, which requires Neumann boundary condition at both ends, $\phi'(0)=\phi'(1)=0$.

First, the boundary conditions accommodate a massless mode with constant profile, which can also be read off directly from (\ref{phi-eqn-2}),
$\phi_{0}(y)=c$, where $c$ is a constant. Because the classical $U(1)$ R-symmetry of the gauge theory is anomalous and reduced to a discrete $Z_{2N}$ symmetry in the quantum theory, its breaking to $Z_2$ by gaugino condensation does not lead to a massless Goldstone boson. Therefore, we avoid this mode by setting $c=0$, since it is constant and nondynamical.
To find the massive modes, first we impose $\phi'(1)=0$  and  (\ref{phi-sol-1}) reduces to
\be
\phi(y)=\frac{C}{\sqrt{y}} \left(M_{\frac{\lambda}{4},\frac{1}{4}}(\lambda
   y^2)+ \frac{(\lambda+3)
   M_{\frac{\lambda}{4}+1,\frac{1}{4}}(\lambda)+(\lambda-1)
   M_{\frac{\lambda}{4},\frac{1}{4}}(\lambda)}
    {4 W_{\frac{\lambda}{4}+1,\frac{1}{4}}(\lambda)-(\lambda-1)
   W_{\frac{\lambda}{4},\frac{1}{4}}(\lambda)}\, W_{\frac{\lambda}{4},\frac{1}{4}}(\lambda y^2)\right).\label{phi-sol-2n}
\ee

\begin{figure}[t]
\centering
\subfloat[$n=1$.]{\label{fig:phi1nn}\includegraphics[width=0.4\textwidth]
{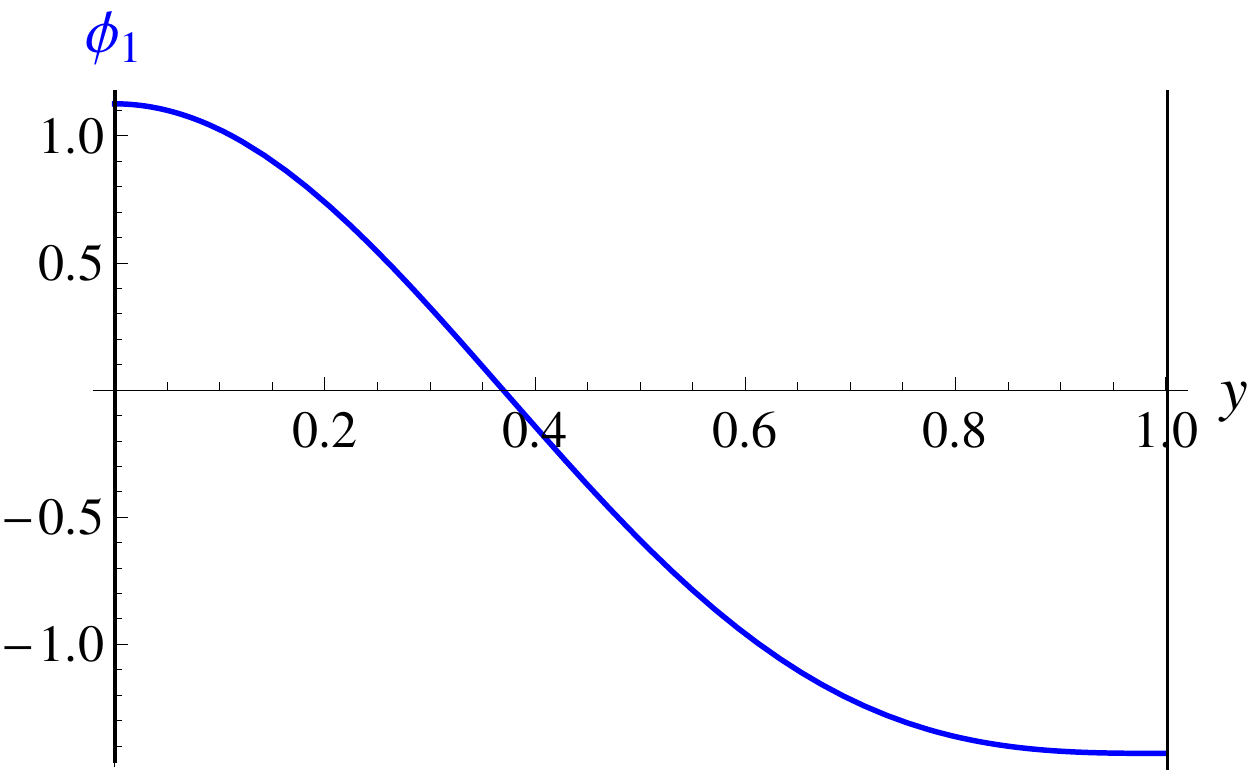}}
\hspace{20mm}
\subfloat[$n=2$.]{\label{fig:phi2nn}\includegraphics[width=0.4\textwidth]
{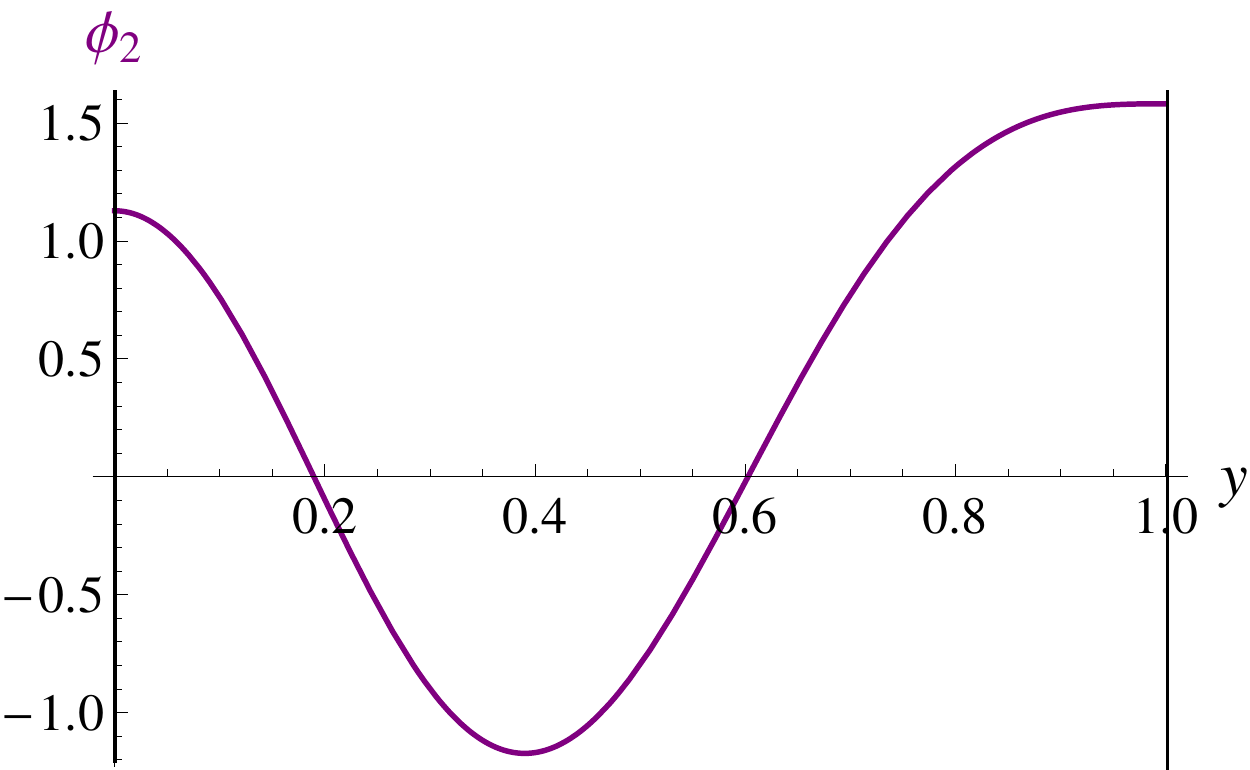}}
\hspace{20mm}
\subfloat[$n=3$.]{\label{fig:phi3nn}\includegraphics[width=0.4\textwidth]
{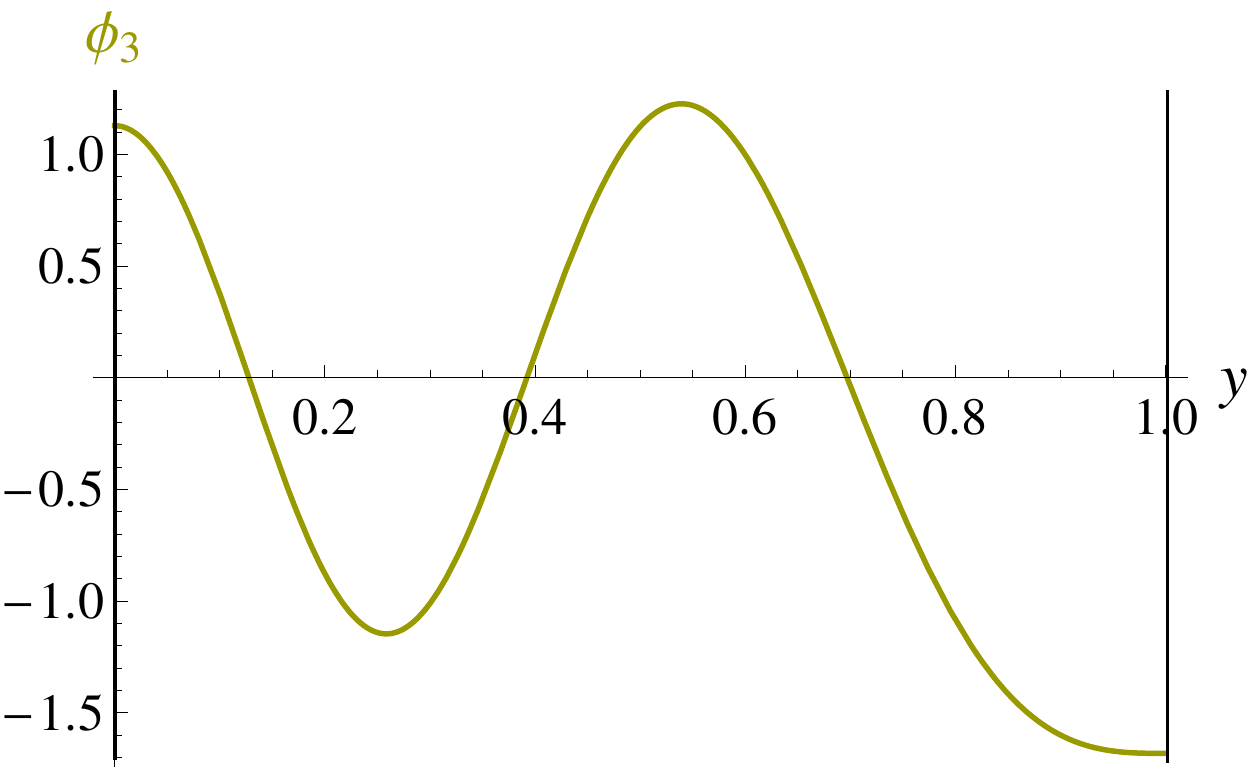}}
\hspace{20mm}
\subfloat[$n=4$.]{\label{fig:phi4nn}\includegraphics[width=0.4\textwidth]
{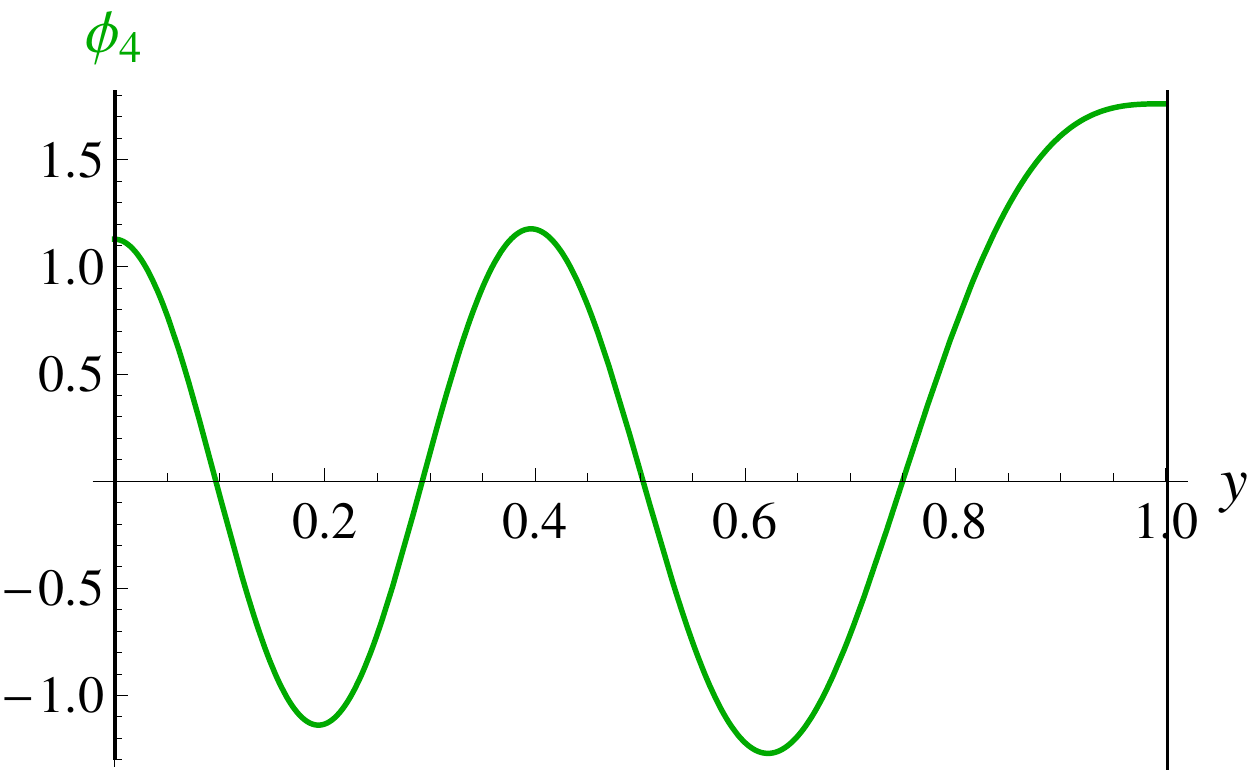}}
\caption{The first few modes $\phi_n(y)$ plotted using the expression given by (\ref{phi-sol-2n}) with the roots $\lambda_n$ obtained from (\ref{lambdaroots-1}) for $\lambda$.}\label{figs:phimodes-1}
\end{figure}

We then impose $\phi'(0)=0$  to obtain discrete values of $\lambda_n$ as roots of the function $f(\lambda)$ given by (\ref{lambdaroots-1}), and the corresponding modes $\phi_n(y)$ are obtained using $\lambda_n$ for $\lambda$ in (\ref{phi-sol-2n}).
The first few numerical values of $\lambda_n$ are given in Table \ref{tab:lambda}.
\begin{table}[h]
\bc
\begin{tabular}{|c|c|c|c|c|c|c|c|c|c|c|c|}
  \hline
  % after \\: \hline or \cline{col1-col2} \cline{col3-col4} ...
  $\lambda_1$ &$\lambda_2$ &$\lambda_3$ &$\lambda_4$ &$\lambda_5$&$\lambda_6$ &$\lambda_7$ &$\lambda_8$ &$\lambda_9$ &$\lambda_{10}$\\
  \hline
   4.287 &8.304 &12.311 &16.315& 20.317& 24.319 &28.320 &32.321 &36.322& 40.323 \\
  \hline
\end{tabular}
\caption{The first few roots of $f(\lambda)$. The glueball masses are given by (\ref{mn-1}).}
\label{tab:lambda}
\ec
\end{table}
The roots are approximately given by
$\lambda_n\approx 4.3\, +\,4(n-1),\, n=1,2,\cdots$. One way to obtain (\ref{lambdaroots-1}) is to make a series expansion of (\ref{phi-sol-2n}) in $y$, set the coefficient of the linear term to zero, and use the recursion relation $W_{k,\mu}(\lambda)=\frac{\Gamma(-2\mu)}{\Gamma(1/2-\mu-k)}M_{k,\mu}(\lambda)
+\frac{\Gamma(2\mu)}{\Gamma(1/2+\mu-k)}M_{k,-\mu}(\lambda)$ that is valid for non-integer $2\mu$, which is the case here.
The first few modes, normalized as $\int_0^{1}dy\,\phi_n(y)^2=1$, are plotted in Figure \ref{figs:phimodes-1} using (\ref{phi-sol-2n}) with the numerical values for $\lambda_n$ obtained from the roots of $f(\lambda)$.

The glueball wavefunctions are nonzero and have largest magnitudes, and increasingly so for larger $n$, at the location of a four-dimensional observer at the UV boundary. On the other hand, modes of fluctuations in the angular directions near the UV boundary, where the radius of the transverse internal space gets smaller and spacetime becomes four-dimensional, are much heavier.

\begin{figure}[t]
\centering
{\includegraphics[width=0.7\textwidth]{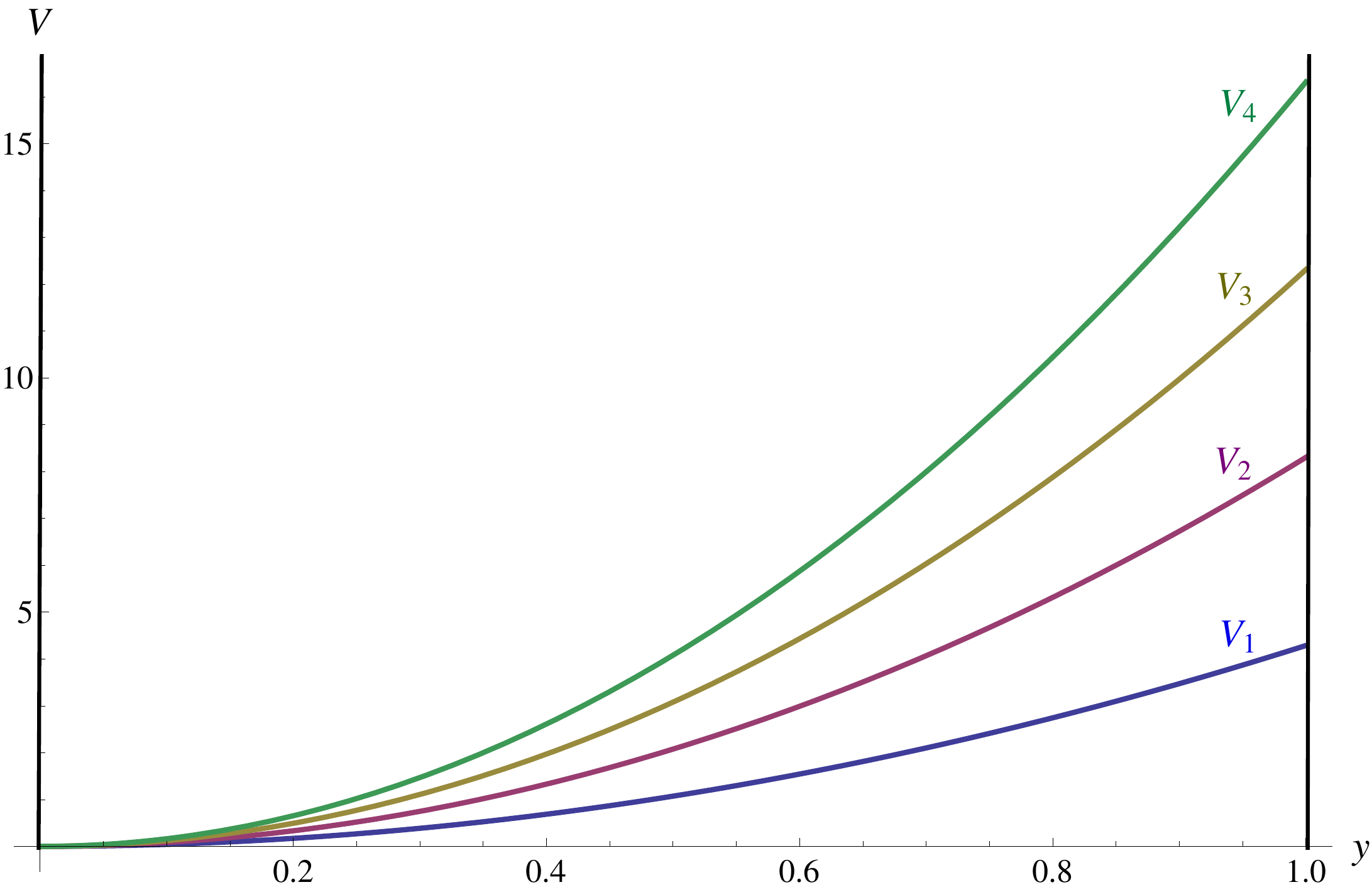}}
\caption{The first few $V_n(y)$. The glueball masses are given by (\ref{mn-1}),  where $\lambda_n=V_n(1)$ are the roots of (\ref{lambdaroots-1}).}\label{figs:V03-1a}
\end{figure}

The equation of motion of $\phi(y)$ given by (\ref{phi-eqn-2}) can be interpreted as a Shr\"{o}dinger equation of the form $\phi''(y)-\lambda V(y)\phi(y)+\lambda E\phi(y)=0$ with $V(y)=\lambda y^2$ and $E=\lambda=V(1)$. The different values of $\lambda_n$ can then be viewed as energies corresponding to different modes $\phi_n(y)$ for potentials $V_n(y)$ with depths which increase with increasing $\lambda_n$.
Plots of the first few $V_n(y)$ are shown in Figure \ref{figs:V03-1a}.

\section {Comparison to lattice QCD data\label{sec:comp-lattica-data}}

The ratios of the masses are independent of the parameters of the theory, $\frac{m_{n+1}}{m_n}=\frac{\lambda_{n+1}}{\lambda_n}$, and serve as predictions.
A lattice computation for the lowest three $0^{++}$ glueball masses in non-supersymmetric $SU(\infty)$ QCD is given in \cite{Lucini:2010nv}, with the states denoted by $A_1^{++}$, $A_1^{++\star}$, and $A_1^{++\star\star}$. This allows us to compare two independent mass ratios as shown in Table \ref{tab:compare}. Both ratios agree to within five percent.
\begin{table}[h]
\bc
\begin{tabular}{|c||c|c|c|c|c|}
  \hline
  % after \\: \hline or \cline{col1-col2} \cline{col3-col4} ...
   & Supergravity & Lattice \\
  \hline\hline
   $\frac{m_{2}}{m_1}$ &$\frac{8.304}{4.287}= 1.937$ &$\frac{1.456}{0.778}= 1.871$ \\
   \hline
   $\frac{m_{3}}{m_2}$& $\frac{12.311}{8.304}= 1.483$ &$\frac{2.061}{1.456}= 1.416$ \\
  \hline
\end{tabular}
\caption{Comparison of mass ratios between our supergravity theory and the non-supersymmetric $SU(\infty)$ lattice QCD data in \cite{Lucini:2010nv}.}
\label{tab:compare}
\ec
\end{table}

\section{Conclusions\label{sec:concl}}

The supergravity background that we argued in \cite{Hailu:2011pn,Hailu:2011kp} to correspond to pure $\mathcal{N}=1$ $SU(N)$ gauge theory in four dimensions in the large $N$ limit in the IR produces mass spectrum of $0^{++}$ glueballs that is qualitatively consistent with expectations from the gauge theory side and
quantitatively in striking agreement with available data from non-supersymmetric large $N$ lattice QCD.

Because the background geometry is compact, the discreteness of the spectrum follows with appropriate boundary conditions imposed at the IR and the UV boundaries.
The magnitudes of the masses become smaller as the 't Hooft coupling at the UV boundary gets smaller, consistent with expectation from the gauge theory side where a smaller 't Hooft coupling corresponds to smaller effective gauge coupling in the large $N$ limit. From the gravity theory point of view, a smaller 't Hooft coupling at the UV boundary corresponds to a larger size of the compact background and smaller curvature which gives modes with smaller masses. A four dimensional observer at the UV boundary measures a 't Hooft coupling of magnitude $g_sN$ that can be chosen appropriately. Furthermore, Kaluza-Klein excitations in the angular directions near the UV boundary are much heavier and a  four-dimensional observer sees only the radial excitations.

The ratios of the masses are independent of the parameters of the theory and the interval between consecutive masses is nearly constant and approximately $4(\frac{g_sN}{2\pi r_s})$ which can be chosen appropriately to produce absolute magnitudes of the masses.
Comparison of the mass ratios to recent non-supersymmetric large $N$ lattice QCD data available in \cite{Lucini:2010nv} for the first three masses shows agreement of the two independent ratios to within five percent. The supergravity expressions for the masses and wavefunctions in this note include the higher mode $0^{++}$ glueballs and additional lattice data would be quite interesting for further comparison.

It would also be interesting to study metric and flux fluctuations on the background and calculate the spectrum of other $J^{PC}$ glueballs and to investigate their relations to the $0^{++}$ glueballs and see if they fall on Regge trajectories.

We would like to re-emphasize that the supergravity background presented in \cite{Hailu:2011pn,Hailu:2011kp} and briefly summarized in Section \ref{sec:sugrab} has a number of remarkable features. It reproduces the renormalization group flow and the pattern of chiral symmetry breaking of the gauge theory. It produces linear confinement of quarks in four dimensions within relativistic quantum theory of ten-dimensional superstrings. The background is compact and the curvature is smallest in the IR where the gauge theory is strongly coupled and a dual gravity description is useful. The radius of the transverse internal space at the IR boundary is set by the nonperturbative scale of the gauge theory and is independent of the 't Hooft coupling. Smaller 't Hooft coupling at the UV boundary corresponds to smaller curvature of the background. The background preserves supersymmetry and is stable.

The linear confinement of quarks the background produces together with the agreement of the ratios of the $0^{++}$ glueball masses obtained in this note with available non-supersymmetric large $N$ lattice QCD data means to us that $\mathcal{N}=1$ $SU(N)$ gauge theory has similar mass spectra as in QCD and the supergravity theory can be used for exploring additional nonperturbative phenomena in QCD.

\newpage
\providecommand{\href}[2]{#2}\begingroup\raggedright\endgroup

\end{document}